\documentclass[10pt, conference]{IEEEtran}
\IEEEoverridecommandlockouts
\usepackage{algorithm}
\usepackage{cite}
\usepackage{amsmath,amssymb,amsfonts}
\usepackage{algorithmic}
\usepackage{graphicx}
\usepackage{textcomp}
\usepackage{xcolor}
\usepackage{epsfig} 
\usepackage{mathptmx} 
\usepackage{times} 
\usepackage{float}
\usepackage{romannum}
\usepackage{setspace}
\usepackage{amsthm}
\usepackage{verbatim}

\usepackage{multirow}
\usepackage{caption}
\usepackage{subcaption}
\usepackage{txfonts}
\usepackage[cal=cm, scr=rsfs]{mathalpha}

\usepackage[letterpaper, left=0.61in,right=0.61in,bottom=0.986in,top=0.71in]{geometry}

\def\BibTeX{{\rm B\kern-.05em{\sc i\kern-.025em b}\kern-.08em
    T\kern-.1667em\lower.7ex\hbox{E}\kern-.125emX}}

\begin{document}

\title{Channel Gain Map Reconstruction Based on Virtual Scatterer Model} 

\author{\IEEEauthorblockN{He Sun\IEEEauthorrefmark{1}\IEEEauthorrefmark{2}, 
                           Lipeng Zhu\IEEEauthorrefmark{1}\IEEEauthorrefmark{3}, 
                           Jie Xu\IEEEauthorrefmark{4}, 
                           and Rui Zhang\IEEEauthorrefmark{1}}
\IEEEauthorblockA{\IEEEauthorrefmark{1}Department of Electrical and Computer Engineering, National University of Singapore, Singapore 117583.} 
\IEEEauthorblockA{\IEEEauthorrefmark{2}School of Electronic and Information Engineering, Beihang University, Beijing 100191, China.} 
\IEEEauthorblockA{\IEEEauthorrefmark{3}State Key Laboratory of CNS/ATM, Beijing Institute of Technology, Beijing 100081, China.} 
\IEEEauthorblockA{\IEEEauthorrefmark{4}School of Science and Engineering (SSE), the Shenzhen Future Network of Intelligence Institute (FNii-Shenzhen), \\ 
Guangdong Provincial Key Laboratory of Future Networks of Intelligence, \\  The Chinese University of Hong Kong (Shenzhen), Guangdong 518172, China.} 
E-mail: {sunhe1710@buaa.edu.cn, zhulp@bit.edu.cn, xujie@cuhk.edu.cn, elezhang@nus.edu.sg}                    }
\maketitle

\begin{abstract}
This paper proposes an efficient method for modeling and reconstructing the channel gain map (CGM) based on virtual scatterers. Specifically, we develop a virtual scatterer model to characterize the channel power gain distribution in three-dimensional (3D) space, by capturing the multi-path propagation environment structure and exploiting the angular-domain spatial correlation of scatterer response. In this model, the CGM is represented as a function over a set of tunable parameters for virtual scatterers, including their number, positions, and scatterer response coefficients (SRCs), which can be estimated from a limited number of channel power gain measurements at a given set of locations within the region of interest. This new representation offers a flexible and scalable modeling framework for efficient and accurate CGM reconstruction. Furthermore, we propose a progressive estimation algorithm to acquire the scatterers' parameters. In this algorithm, we gradually increase the number of virtual scatterers to balance the computational complexity and estimation accuracy. In addition, by exploiting the spatial correlation of scatterer response, we propose a Gaussian process regression (GPR)-based inference method to predict the SRCs that cannot be directly estimated. Finally, ray-tracing-based simulation results under realistic physical environments validate the effectiveness of the proposed method, demonstrating that it achieves higher reconstruction accuracy compared to conventional CGM estimation approaches.
\end{abstract}


\section{Introduction}
The global coverage, high user mobility, and dense node deployment envisioned in future wireless communication systems impose stringent requirements on accurate and efficient channel information acquisition\cite{WangChengXiangEIT}. Traditional channel acquisition methods typically rely on extensive channel measurements or pilot-based channel estimation, while the burden for implementing them becomes heavier for higher-dimensional channels due to increasingly larger-scale antenna arrays and wider bandwidth in future wireless networks\cite{WangChengXiangEIT,ZengYT,CaoSurvey}. To address this issue, channel knowledge map (CKM) has been introduced\cite{ZengYT}, whose construction methods can efficiently generate channel parameters at unmeasured locations by exploiting environment information, thereby reducing or even eliminating the need for real-time channel estimation. 

As a key instance of CKMs, the channel gain maps (CGMs) characterize the spatial variation of large-scale fading across a target region, providing essential information for resource allocation, low-altitude flight trajectory planning, coverage enhancement, and so on\cite{ZengYT,RMSMAP,TWCIRS1}. CGM construction is a foundational problem that underpins its practical applications\cite{WangChengXiangEIT}. To tackle this problem, a variety of algorithms have been proposed. Based on their underlying modeling methodology, the CGM construction methods can be generally categorized into two classes: (i) electromagnetic (EM) model-driven channel derivation, and (ii) data-driven CGM construction approaches. Specifically, the former approaches integrate site geometry, EM material parameters, and antenna radiation patterns into EM propagation models to accurately compute the channel gain between transceivers, such as EM information theory (EIT)-based approaches\cite{WangChengXiangEIT}, and ray-tracing-based approaches\cite{RTS0}. However, these EM model-driven approaches require detailed physical environment information, which is highly challenging to acquire, especially in dense scattering environments or wide-area outdoor regions. Alternatively, the data-driven approaches directly model the channel parameters from empirical observations and estimate the model parameters for CGM construction based on channel measurements, without explicitly accounting for the effects of scatterers in EM wave propagation. Typical techniques include segmented-channel model-based radio mapping algorithms\cite{SGE1}, expectation maximization estimation methods\cite{EM}, Kriging interpolation\cite{Kriging}, deep neural network (DNN)-based methods\cite{RMSDNN}, and wireless radiance field methods like bidirectional wireless Gaussian splatting (BiWGS) \cite{XUBIWGS}. Although these methods avoid the need for detailed physical environment information, they typically require sufficient channel measurement data for accurate CGM construction.

Since future communication networks encounter complex scattering in heterogeneous propagation environments such as urban areas, the limited region accessibility and sensing capability make it difficult and costly to acquire large-scale channel measurements and/or detailed EM parameters of scatterers. To cope with this difficulty, a new scatterer model was proposed in \cite{CGM01} for physics-guided CGM estimation based on a small set of channel power gain measurements collected at designated locations in the area of interest. The scatterer model is designed based on the physical structures of the multipath channel, which focuses on modeling the signal response rather than explicitly simulating the detailed interactions between the EM waves and environments, thus enabling a lightweight and digital characterization of the CGM. However, this approach ignores the angular-domain correlation of the scatterer response and therefore requires at least one measurement in each angular sector surrounding each physical scatterer, leading to an explosion of model parameters and an excessive measurement overhead in dense scattering environments.

 To overcome the above issues, we propose in this paper a general three-dimensional (3D) {\it virtual scatterer}-based scatterer-centric environment modeling approach for physically grounded CGM representation and efficient CGM estimation. Different from \cite{CGM01}, the number and positions of virtual scatterers are modeled as tunable parameters, which are estimated based on channel power gain measurements, thus reducing the reliance on information of physical scatterers. To capture the angular correlation of scatterer response, the scatterer response coefficient (SRC) of each virtual scatterer is further modeled by a Gaussian random process indexed by the angle of departure (AoD) of the signal leaving the scatterer. As a result, the SRC corresponding to a given AoD can be inferred based on those associated with other AoDs, thus avoiding the need for exhaustive directional coverage in channel measurements. An efficient progressive estimation algorithm is developed for CGM construction by using the scalable virtual scatterer model, which remains effective even in dense scattering environments. Moreover, a Gaussian process regression (GPR)-based inference method is designed to predict the SRCs corresponding to unmeasured locations, enabling a complete CGM reconstruction. Simulation results using a practical 3D ray-tracing channel model verify that the proposed algorithm can reconstruct the CGM based on only a small number of channel power gain measurements, and it achieves higher estimation accuracy than various benchmarks.

\emph{Notations}: $\mathbb{R}^{n \times m}$ denotes the set of $n \times m$ real matrices. $[\boldsymbol{F}]_{ i , j } $ denotes the element in the $i$-th row and the $j$-th column of matrix $\boldsymbol{ F }$. $ \mid \cdot \mid $ denotes the cardinality of a set or the amplitude of a complex number. $\parallel\cdot\parallel$ denotes the Euclidean norm of a vector. The $n$-dimensional Gaussian distribution with mean vector $\boldsymbol\mu_{ n } \in \mathbb{ R } ^ { n \times 1 }$ and covariance matrix $\mathbf V_{ n } \in \mathbb{R}^{ n \times n }$ is denoted by $\mathcal N(\boldsymbol\mu_{ n } , \mathbf V_{ n } ) $. The modulo operation $\text{mod}( a , b )$ returns the remainder when the integer $a$ is divided by a positive integer $b$.

\section{CGM Characterization Based on Virtual Scatterer Model}\label{sec002}

\subsection{Virtual Scatterer-Based Multipath Channel Model}
\begin{figure}[t]
  \centering
  \begin{subfigure}[b]{0.23\textwidth}
    \centering
    \includegraphics[width=\textwidth]{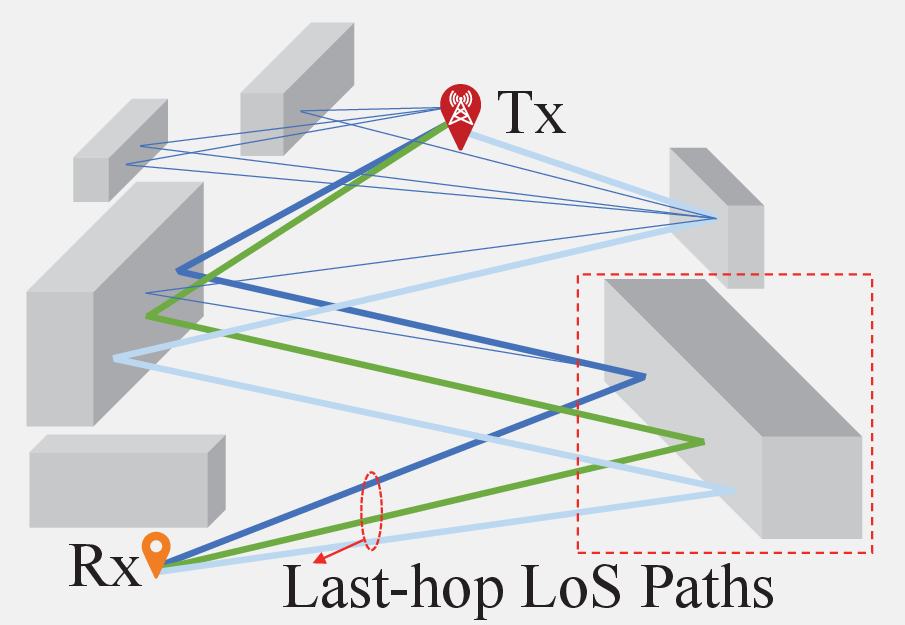}
    \caption{}
    \end{subfigure}
  \hfill
  \begin{subfigure}[b]{0.23\textwidth}
    \centering
    \includegraphics[width=\textwidth]{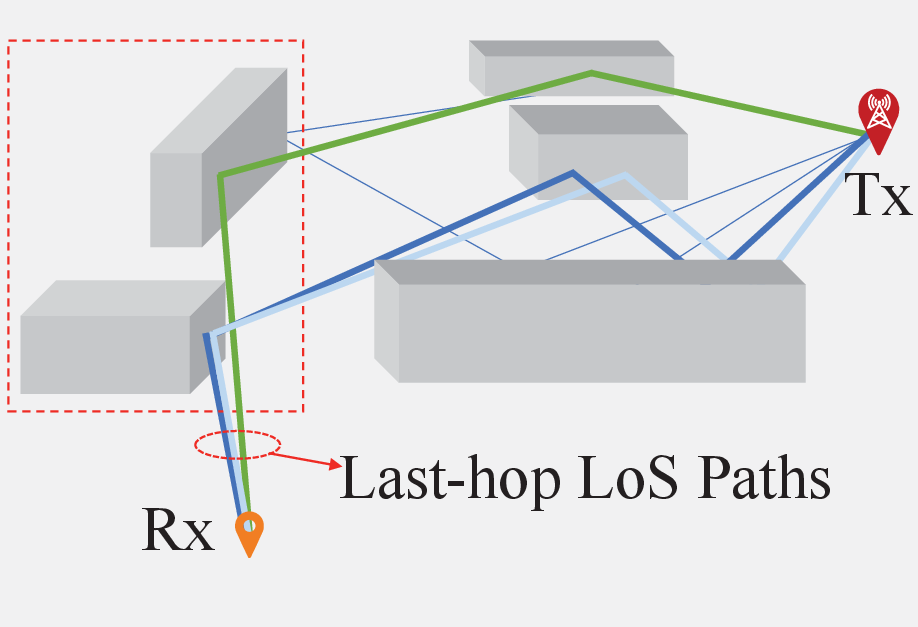}
    \caption{ }
  \end{subfigure} 
  \caption{Multipath channel generated by ray tracing} \vspace{-16.6pt} 
  \label{fig002}
\end{figure}

In a 3D region containing numerous scatterers, signals from the transmitter (Tx) propagate by reflection, refraction, and diffraction at densely distributed scatterers, leading to a complex multipath propagation environment. Accurate description of the signal propagation paths among scatterers is challenging due to the infinite number of interactions between the signals and the environment, which renders the channel characterization highly complex and computationally prohibitive. However, for the multipath channel observed at each receiver (Rx) location, each multipath component terminates at the Rx location with a line-of-sight (LoS) signal path, which originates either from the Tx or from one of the scatterers, as shown in Fig. \ref{fig002}. Instead of fully tracing the physical interactions between the signals and the environment, the scatterer model-based channel characterization method uses only the last-hop LoS paths from scatterers and/or the Tx to represent the channel response at each Rx location, thereby ensuring consistency with physical propagation conditions while reducing the modeling complexity. In particular, physical scatterers that contribute significantly to the total received signal power or have a major impact on the structure of the propagation environment are referred to as \emph{dominant scatterers}, such as buildings, trees, or other large-scale infrastructures\cite{CGM01}. To accurately characterize the signal response of dominant scatterers, we propose in this paper a virtual scatterer model by representing physical dominant scatterers as a set of \emph{virtual scatterers}. 

As shown in Fig. \ref{fig002}(a), the last-hop LoS paths originate from different parts of a large physical scatterer, which motivates representing this physical scatterer as multiple virtual scatterers for accurate channel characterization. In contrast, as shown in Fig. \ref{fig002}(b), two closely located physical scatterers with similar propagation characteristics, such as similar LoS visibility and signal propagation direction from scatterers to remote Rx locations, can be jointly represented by a single virtual scatterer to simplify the model. For clarity and modeling flexibility, we treat the number of virtual scatterers as a tunable parameter\footnote{The method for determining $N$ will be presented in Section \ref{sec003}.} denoted by $N$. Let $\boldsymbol{s}_n \in \mathbb{R}^{ 3 \times 1}$ denote the 3D coordinate of the $n$-th virtual scatterer, $ n = 1 , \cdots , N $. In particular, $\boldsymbol{s}_0 \in \mathbb{R}^{ 3 \times 1}$ denotes the 3D coordinate of the Tx.

\begin{figure}[t]
  \centering
  {\includegraphics[width=0.369\textwidth]{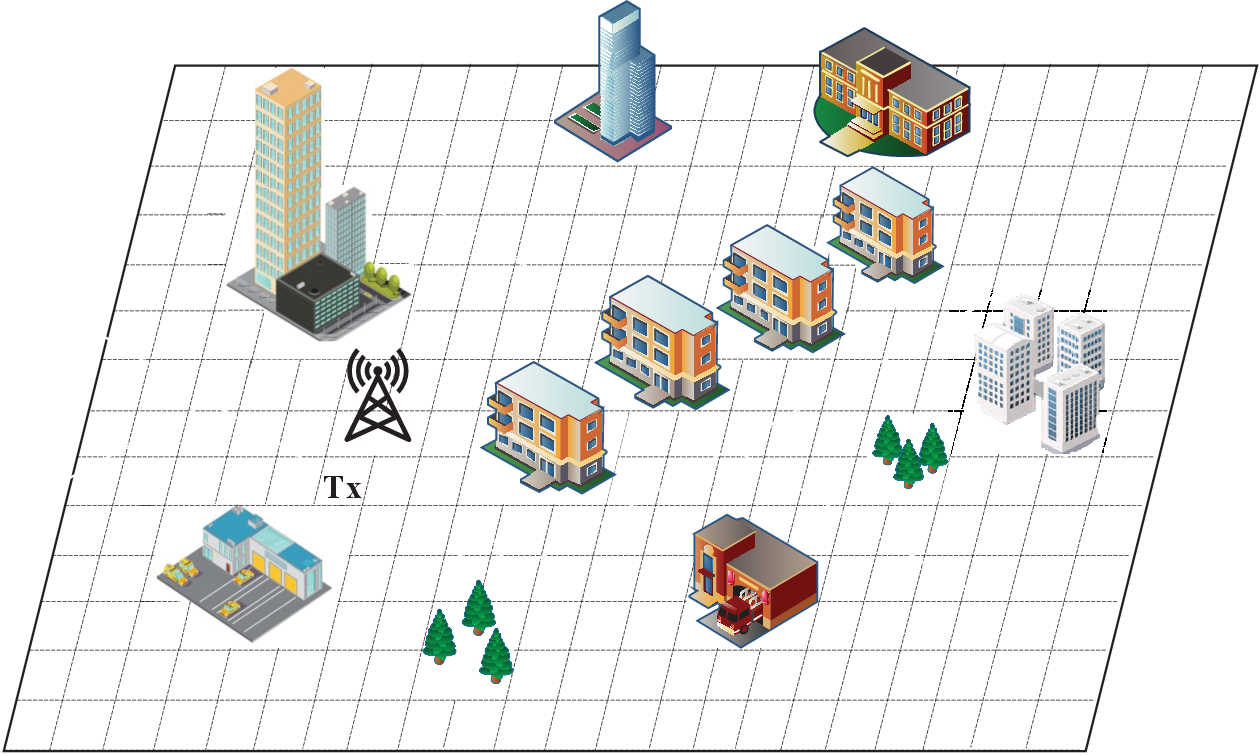}}
  \caption{Uniform square-grid partition of the region.}
  \vspace{-16pt}
\label{Fig0003}
\end{figure}

 For any location within a small-size local area in the region, the set of dominant physical scatterers, and thus their virtual-scatterer representation, are similar, yielding a locally homogeneous propagation environment in which the multipath components at all locations in the local area are deterministic\cite{CGM01}. To facilitate CGM characterization by exploiting this fact, the plane of interest can be partitioned into a set of equally spaced square grids\footnote{Without loss of generality, we consider CGM reconstruction on a 2D plane in the 3D region. However, the 2D CGM can be extended to a 3D map by using multiple 2D CGMs on parallel planes at different heights above the ground.}, as shown in Fig. \ref{Fig0003}. Let $ { \cal{ I } } \triangleq \{ 1, 2, \cdots, I\}$ denote the set of indices for grids not occupied (either partially or fully) by the Tx or any physical scatterer, and $I$ represent the total number of such grids. Since most grids are far from the Tx in an outdoor region, the propagation environment is considered in the far field from the Tx, where the path power gain and the angle of arrival (AoA) of each multipath component remain constant across different locations within the same grid\cite{CGM01}. Let $\beta_{n,i}$ denote the path power gain of the LoS path from the Tx or a virtual scatterer $n$ to grid $i$. By setting the grid size to be the order of several tens of wavelengths, the average channel power gain over a grid is equal to the linear sum of the path power gains of all multipath components arriving at that grid\cite{CGM01,MAS}, i.e.,
\begin{equation}\label{Eqs001}
  q_i = \sum_{n \in {\cal{V}}_i} \beta_{n,i}, \ \ i \in \mathcal{ I } ,
\end{equation}
where ${\cal{V}}_i$ is a set consisting of virtual scatterers that have an LoS link to grid $i$. The proof of (\ref{Eqs001}) is similar to that of Lemma 1 presented in \cite{CGM01}, and thus is omitted here. Specifically, since different multipath components arrive at the Rx location from distinct AoAs, their phase differences vary rapidly in space, thus resulting in channel power gain fluctuations occurring over distances of a wavelength. However, as the grid size is much larger than the wavelength, such fluctuations only change the spatial distribution of the channel power, while the average channel power within each grid remains unchanged.

To maintain consistency with the physical geometry of multipath channels, the LoS visibility between a virtual scatterer and a grid is determined by checking whether the associated physical scatterer (to be specified in Section \ref{sec00302}) has an LoS path to that grid. Note that the existence of an LoS link between a physical scatterer and a grid can be determined according to the spatial layout and the geometry of physical scatterers, which is readily available from publicly accessible open-source databases such as OpenStreetMap and Natural Earth.\footnote{The databases are available online: https://www.openstreetmap.org, and https://www.naturalearthdata.com.} 

\subsection{CGM Characterization Based on Virtual Scatterer Model}\label{Sec00202}

Based on (\ref{Eqs001}), the average channel power gain within each grid is determined by the power gain $\beta_{n ,i}$ of each LoS path component. To characterize the path power gains $ \beta _{ n , i } $'s, a 3D virtual scatterer model is proposed in the following.

Note that the path power gain from a virtual scatterer to a specific grid can be expressed as the product of the SRC and the propagation power gain along the LoS path\cite{CGM01}. Let $ \tau _{ n , i } $ and $ g _{ n , i } $ denote the SRC of virtual scatterer $ n $ towards the direction of grid $ i $ and the propagation power gain of the LoS path between virtual scatterer $ n $ and grid $ i $, respectively. The path power gain of the LoS path from virtual scatterer $ n $ to grid $ i $ is given by 
\begin{equation}\label{Eqs002}
  \beta _{ n , i } = \tau_{ n , i } g _{ n , i } , \ \ {n \in {\cal{V}}_i}  .
\end{equation}
To characterize the propagation power gain $ g _{ n , i } $, let $\boldsymbol{ s }_{ n }$ and $\boldsymbol{c}_i$ denote the coordinates of virtual scatterer $n$ and the center of grid $i$ in the 3D Cartesian coordinate system, respectively. The propagation power gain is given by
\begin{equation}\label{Eqs003}
  g_{n,i} = \frac{\beta_0}{ || {\boldsymbol{ s }}_{ n } - \boldsymbol{ c } _i || ^{ \alpha }}, \ \  {n \in {\cal{V}}_i} ,
\end{equation}
where $\beta_0$ and $\alpha$ denote the reference path power gain at a reference distance of 1 meter (m) and the path loss exponent, respectively. For simplicity, we assume that all the LoS paths in the region have the same reference path power gain and path loss exponent, and they are known {\it a prior}.

The SRC characterizes the ratio of the signal power reflected/refracted by a scatterer to the transmit power of the Tx. As EM waves impinging on the same scatterer from nearby directions experience similar path lengths, reflection coefficients, and scattering geometries, the SRCs observed at different but close directions exhibit angular correlation. To characterize such angular correlation, the SRCs of each scatterer can be modeled as correlated Gaussian random variables that depend solely on the AoD of the LoS path from virtual scatterer $n$ to grid $i$. Let $\boldsymbol{\varphi}(n,i) \in \mathbb{R}^{2\times 1} $ denote the azimuth and elevation AoDs of the LoS path from scatterer $n$ to grid $i$. The SRC of virtual scatterer $ n $ towards the direction of grid $ i $ can be expressed as a function of the AoD of LoS path between scatterer $n$ and grid $i$, i.e., $ \tau_n( { { \boldsymbol{ \varphi } } } ( n , i ) ) $. To simplify the scatterer model, the AoDs are uniformly divided into $M$ sectors in a 3D coordinate system. Let $ \{ \boldsymbol{\varphi}_1 , \boldsymbol{\varphi} _2 , \ldots, \boldsymbol{\varphi}_{ M } \}$ denote the set of AoD sectors. As such, the SRCs of each virtual scatterer $n$ can be modeled as $M$ correlated Gaussian random variables, i.e., $\boldsymbol{\tau} _n = \big[ \tau_n( \boldsymbol{\varphi}_1 ), \tau_n( \boldsymbol{\varphi} _2 ), \ldots, \tau_n( \boldsymbol{\varphi}_{ M } ) \big ]^{ T } \in \mathbb{ R } ^{ M \times 1 }$, and the SRC of virtual scatterer $ n $ towards the direction of grid $ i $ can be computed by $ \tau_n( { { \boldsymbol{ \varphi } } } ( n , i ) ) = \tau_n( \boldsymbol{\varphi}_{ m } )$, where the angles $ { { \boldsymbol{ \varphi } } } ( n , i ) $ lie in the $m$-th AoD sector $\boldsymbol{\varphi}_{ m }$, $m=1, 2 , \cdots , M$. By substituting (\ref{Eqs002}) and (\ref{Eqs003}) into (\ref{Eqs001}), the CGM can be denoted by
\begin{equation}\label{Eqs005}
  q_i = \sum_{n \in {\cal{V}}_i} \beta_{n,i} = \sum_{n \in {\cal{V}}_i} \frac{\beta_0}{|| \boldsymbol{s}_{n} - \boldsymbol{c}_i ||^{ \alpha }}\tau _n ( { {\boldsymbol{\varphi}} } (n,i)), \ \ \ i \in \cal{ I }.
\end{equation}
Based on (\ref{Eqs005}), the CGM is represented by a function of the number, positions, and SRCs of virtual scatterers, collectively termed as \emph{scatterer parameters}. In practice, each $q_i$ in (\ref{Eqs005}) can be obtained by resolving the multipath channel power gain $\beta_{n,i}$'s of the multipath components using a multi-antenna Rx, or alternatively by measuring the average received signal power at multiple positions within each grid using a mobile sensor. However, such channel measurements should be conducted
at every grid $ i \in \cal{ I } $, which imposes prohibitive overhead in large-scale regions. To address this difficulty, the CGM can be constructed by specifying the scatterer parameters. This reformulation converts CGM construction into a scatterer parameter estimation problem, as detailed in the next section.

\section{ CGM Reconstruction Based on Progressive Estimation }\label{sec003}

In this section, we present a CGM reconstruction framework, where the number and positions of virtual scatterers and their SRCs over measured directions are first estimated based on average channel power measurements. Moreover, a GPR-based SRC inference method is designed to predict the SRCs over unmeasured directions by exploiting the angular-domain correlation of scatterer response. Finally, the entire CGM can be constructed based on the estimated scatterer parameters.

\subsection{Scatterer Parameter Estimation Problem Formulation}

Let $ { \cal{ L } } \subset \cal{ I } $ denote the set of indices for arbitrarily selected grids where the channel power measurement is conducted, with $ L \triangleq\left|{ \cal{ L } }\right| $ and $ L < I $. Define $\bar{q}_i$ and $\hat{q}_i$ as the measured average channel power gain within grid $i \in {\cal{ L }}$ and its estimation, respectively, where $\hat{q}_i$ can be computed by (\ref{Eqs005}) based on the estimated scatterer parameters. Let $\tilde{\boldsymbol{ \tau } }$ denote the set of SRCs involved in the estimations of the channel power gains for $L$ measurement grids, and ${ \cal{ S }  } _N  \triangleq \{ {\boldsymbol{ s }}_{ 1 } , {\boldsymbol{ s }}_{ 2 } , \cdots, {\boldsymbol{ s }}_{ N }  \} $ denote the set of coordinates of $N$ virtual scatterers. The scatterer parameter estimation problem can be formulated as
\begin{subequations}
\begin{align}\label{Eqs006}
  \text{(P1):}  & \min_{N, \tilde{\boldsymbol{ \tau } }, { \cal{ S }  } _N } \frac{ 1 }{ L } \sum_{ i \in { \cal{ L } } } \text{U}( \bar{ q }_i , \hat{ q }_i ), \\
   & \text{s.t. } { \boldsymbol{ s } }_{  n } \in { \cal{ R } } , \ \ \forall n \in \{ 1 , 2, \cdots , N \} , \label{eqs00602}
\end{align}\end{subequations}
where $\text{U}( \bar{ q }_i , \hat{ q }_i )$ denotes a non-decreasing function of squared error $ ( \bar{ q }_i - \hat{ q }_i )^2$, such as the mean square error (MSE) or the normalized MSE (NMSE). Constraint (\ref{eqs00602}) confines the position of each virtual scatterer to a predefined 3D space $\cal{ R }$ to mitigate over-fitting. In problem (P1), the number of parameters to be estimated and the form of the objective function vary with the number of virtual scatterers, $N$. This coupling yields a non-convex and variable-dimension optimization problem for which a closed-form solution is intractable. Specifically, increasing $N$ improves the model's degrees of freedom (DoFs) and fitting capacity, while it may lead to a higher computational complexity and the model over-fitting with finite power measurements. To properly balance the estimation accuracy and complexity, we adopt a progressive estimation approach to solve (P1) by incrementally increasing $N$ and refining the solutions until the estimation error converges, as detailed next.

\subsection{Progressive Estimation}\label{sec00302}

Let $N_t$ denote the number of virtual scatterers in the $t$-th progressive iteration, and $N_D$ denote the number of dominant physical scatterers. The progressive estimation starts with $N_1=\varrho$ virtual scatterers, where $\varrho$ denotes the incremental step size of virtual scatterer number in each iteration. Then, we gradually increase the number of virtual scatterers by $\varrho$ and successively refine their positions and SRCs via solving (P1) under a fixed number of virtual scatterers. Specifically, in the $t$-th iteration, the progressive estimation method comprises three stages: initialization of virtual scatterer positions, scatterer position refinement with SRC estimation, and convergence evaluation, as detailed below, respectively.

\subsubsection{Initialization of Virtual Scatterer Positions}

Let ${ \boldsymbol{ s } }_{ n , t }$ denote the initial coordinates of virtual scatterer $n$ in the $t$-th iteration, and $\boldsymbol{ s }_{  n , t }^{ * }$ denote its refined coordinates. Let $\boldsymbol{u}_1, \boldsymbol{u}_2, \cdots, \boldsymbol{u}_{N_D}$ denote the center coordinates of physical scatterers sorted in the descending order of their sizes. For $t=1$, the number of virtual scatterers $N_1 = \varrho$ is set to be smaller than $N_D$ to achieve a lightweight representation of the CGM, and the positions of virtual scatterers are initialized as the geometric centers of the $\varrho$ largest dominant physical scatterers, i.e., $\boldsymbol{s}_{ n , t } = \boldsymbol{u} _{ n } $, $n = 1 , \cdots , \varrho $. For $t > 1$, to reduce the estimation error and accelerate the algorithm's convergence, the positions of $\varrho (t-1)$ virtual scatterers are initialized as the refined ones in the ($t-1$)-th iteration, i.e., $\boldsymbol{s}_{ n , t } = \boldsymbol{s} _{ n , t - 1 } ^{ * }$, $n = 1 , 2 , \cdots , \varrho ( t - 1 ) $. The positions of the remaining $\varrho$ virtual scatterers are initialized based on the locations of the dominant physical scatterers sequentially according to their sizes, i.e., $\boldsymbol{s}_{ n , t } = \boldsymbol{u}_j$, where $j=\text{mod}(n-1,N_D)+1$, and $n = \varrho (t-1) + 1 , \cdots , \varrho t$. 
 
Since the size of a physical scatterer governs the spatial range over which it influences signal propagation, the above initialization method assigns higher priority to larger scatterers by allocating more virtual scatterers to them. Due to the increase in the number of virtual scatterers in each iteration, the set of virtual scatterers with an LoS link to grid $i$, i.e., $ { \cal{ V } }_i $, also changes. This set can be updated based on the LoS visibility between grid $i$ and the physical scatterers corresponding to the newly added virtual scatterers.

\subsubsection{Scatterer Position Refinement and SRC Estimation}

In each progressive iteration, under a fixed number of virtual scatterers $N_t$, the scatterer parameter estimation problem is reduced to (P2), which is given by
\begin{subequations}
\begin{align}
  \text{(P2):}  & \min_{ \tilde{\boldsymbol{ \tau } } _t, { \cal{ S }  } _{ N _t }  } \frac{ 1 }{ L } \sum_{ i \in { \cal{ L } } } \text{U}( \bar{ q } _ i , \hat{ q } _ i ), \\
   & \text{s.t. } { \boldsymbol{ s } }_{  n , t } \in { \cal{ R } } , \ \ \forall n \in \{ 1 , 2, \cdots , N _t \}, \label{eqs00702}
\end{align}\end{subequations}
where $\tilde{\boldsymbol{ \tau } }_t$ denotes the set of SRCs involved in the estimations of the channel power gains in the $t$-th iteration.
Based on (\ref{Eqs005}), virtual scatterer positions are involved in the exponential terms of $ \hat{ q } _ i $ and are coupled with the SRCs, resulting in a high-dimensional and non-convex optimization problem. To improve computational efficiency and accelerate convergence, we propose refining the position of each virtual scatterer successively while keeping those of the others fixed. As such, the refinement of scatterer $n$'s position can be formulated as
\begin{subequations}\label{Eqs007}
\begin{align}
  \text{(P3):}  & \min_{ \tilde{\boldsymbol{ \tau } } _t, { { \boldsymbol{ s } }_{ n , t } }  } \frac{ 1 }{ L } \sum_{ i \in { \cal{ L } } } \text{U}( \bar{ q } _ i , \hat{ q } _ i ), \label{Eqs00701} \\
   & \text{s.t. } { \boldsymbol{ s } }_{  n , t } \in { \cal{ R } } , \ \ \forall n \in \{ 1 , 2, \cdots , N _t \} .
\end{align} \end{subequations}
Note that under fixed virtual scatterer positions $ \boldsymbol{ s }_{ n , t  } $'s, the estimated power gain $ { \hat{ q } } _ i $ in the objective function of (P3) is a linear function of SRCs. Thus, the SRCs can be estimated via solving a least squares (LS) estimation problem under given scatterer positions. To solve problem (P3), we can refine the scatterer positions via the gradient descent algorithm, where the gradients are computed based on the LS solutions of SRCs.

\subsubsection{Convergence Evaluation}

After the $t$-th progressive iteration, the progressive estimation algorithm monitors the convergence to decide whether to proceed to the $(t + 1)$-th progressive iteration with $N_{t+1} = (t+1) \varrho$, or to terminate the progressive estimation process. In particular, the objective function of (P2) is used for convergence evaluation, which is given by
\begin{equation}\label{Eqs014}
  \zeta_t = \frac{ 1 }{ L } \sum_{ i \in { \cal{ L } } } \text{U}\left( \bar{ q }_i , \sum_{n \in { \cal{ V } }_i }  \frac{\beta_0}{|| \boldsymbol{ s }_{ n , t } ^* - \boldsymbol{c}_i || ^{ \alpha } } \tilde{ \tau } _ { n , t } ^{ * } ({ {\boldsymbol{\varphi}} } ( n , i )) \right).
\end{equation}
The progressive estimation can be terminated if $\zeta_t$ has converged or the maximum number of progressive iterations has been reached. In the progressive estimation method, the objective function is monotonically non-increasing, since the model size is progressively increased and the model parameters are optimized via gradient descent. Meanwhile, the objective function values are lower-bounded by zero, so the convergence of the progressive estimation algorithm is guaranteed.


\subsection{CGM Construction}

Due to the limited number of measurements, not all SRCs involved in the virtual-scatterer model-based representation of CGM can be estimated by the above progressive parameter estimation method. To tackle this problem, we propose a GPR-based SRC inference algorithm to predict unestimated SRCs.

 Let $\boldsymbol{ \hat{ \tau } }_n$ denote the subset of entries in $ \boldsymbol{ \tau } _n \in \mathbb{ R } ^{ M \times 1 }$ estimated through the progressive estimation, with $M_n \triangleq \left| \boldsymbol{ \hat{ \tau } }_n \right|$ and $M_n < M $. Based on the virtual scatterer model proposed in Section \ref{Sec00202}, the SRCs follow the multivariate correlated Gaussian distribution. Accordingly, as a subset of $ \boldsymbol{ \tau } _n $, the estimated SRCs follow the $M_n$-dimensional Gaussian distribution, i.e., $\boldsymbol{ \hat{ \tau } }_n \sim \mathcal N(\boldsymbol\mu_{ M_ n },\mathbf V_{ M_ n })$, where $\mathbf V_{ M _ n } \in \mathbb{ R } ^{ M _ n \times M _ n } $ denotes the covariance matrix determined by a Gaussian kernel function given by
\begin{equation}\label{Eqs015}
  [\mathbf{V}_{ M _ n }]_{a,b} = v_n^2 \exp\!\left( - \frac{\|\boldsymbol{ \varphi } _ a - \boldsymbol{ \varphi } _ b \|^2}{2\rho_n^2} \right),
\quad a , b = 1 , \cdots , M _ n ,
\end{equation}
where the kernel function is parameterized by hyperparameters $v_n$ and $\rho_n$, which can be learned by maximizing the log-marginal likelihood of the available estimates $\boldsymbol{ \hat{ \tau } }_n$. 
Specifically, the log-marginal likelihood of $\boldsymbol{ \hat{ \tau } }_n$ is given by \cite{weisberg2005ALR}
\begin{equation}\label{Eqs015e}
 \Upsilon( v _n , \rho _n , \sigma_n )
= -\tfrac{ 1 } { 2 } \boldsymbol{ \hat{ \tau } }_n ^T \mathbf A_{ M_ n } ^{ -1 } \boldsymbol{ \hat{ \tau } }_n
-\tfrac{1} {2} \log| \mathbf A _{ M_ n } |
-\tfrac{ M _n } { 2 } \log( 2 \pi ) ,
\end{equation}
where $\mathbf A_{ M_ n } = \mathbf V_{ M_ n } + \sigma_n ^2 \mathbf I$ is a function of kernel hyperparameters $v_n , \rho_n$, and $\sigma ^2_n$ denotes the standard deviation of the inferred coefficients. By substituting the estimated kernel hyperparameters into (\ref{Eqs015}), the SRC corresponding to any AoD $ \boldsymbol{ \varphi } _ m $ is given by
\begin{equation}\label{Eqs017}
  \hat{\tau}_n(\boldsymbol{ \varphi }_m)
= \mathbf{ v }_{ M _n } ^{ T } \mathbf{ A }_{ M _n } ^{ -1 } \hat{ \boldsymbol{ \tau } } _n ,
\end{equation}
where $ \mathbf{ v } _{ M _n } \in \mathbb{R}^{M _n \times 1}$ denotes the cross-correlation vector between the $ { M _n } $ estimated SRCs of virtual scatterer $n$ and that corresponding to the target AoD $ \boldsymbol{ \varphi } _ m $. By setting $a=m$ in (\ref{Eqs015}), the desired vector $ \mathbf{ v } _{ M _n }$ can be computed by taking the first column of $\mathbf{V}_{ M _ n }$. Finally, the CGM can be constructed by substituting the estimated number and positions of virtual scatterers and their SRCs into (\ref{Eqs005}). The total complexity mainly arises from the progressive estimation of scatterer parameters and the GPR-based SRC inference procedures, and it scales cubically with the number of SRCs estimated for the virtual scatterers.

\section{Simulation Results}\label{sec004}

\begin{figure}[t]
  \centering 
    \includegraphics[width=0.3169\textwidth]{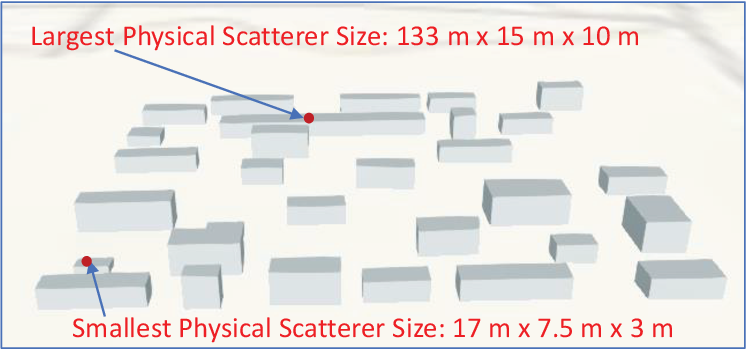}
  \caption{Simulation setup in a 3D environment.}
  \label{Fig004} \vspace{ -9.36pt } 
\end{figure}
\begin{figure}[t]
  \centering
  \begin{subfigure}[b]{0.15\textwidth}
    \centering
    \includegraphics[width=\textwidth]{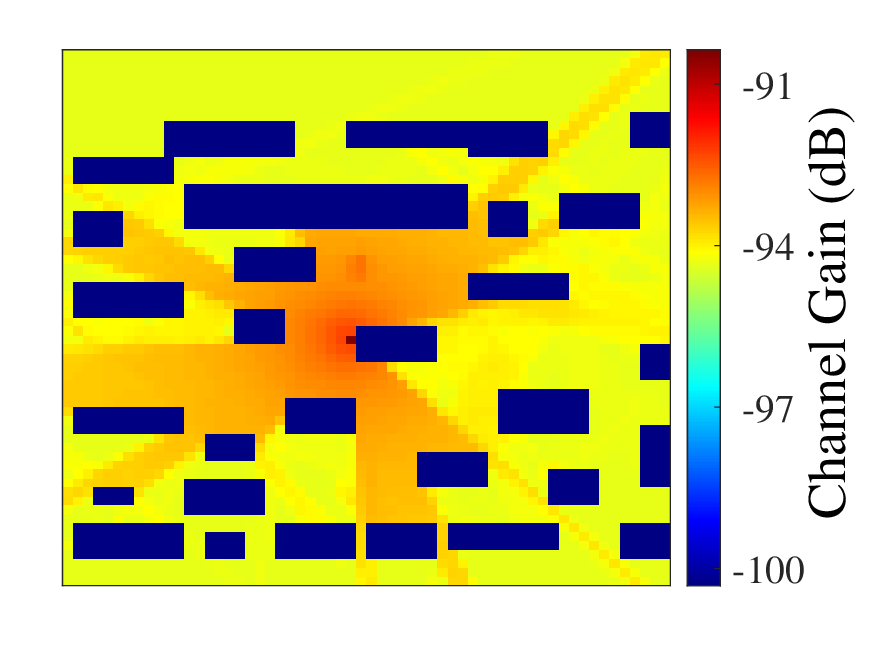}
    \caption{Ground-truth}
  \end{subfigure}
  \hfill
  \begin{subfigure}[b]{0.15\textwidth}
    \centering
    \includegraphics[width=\textwidth]{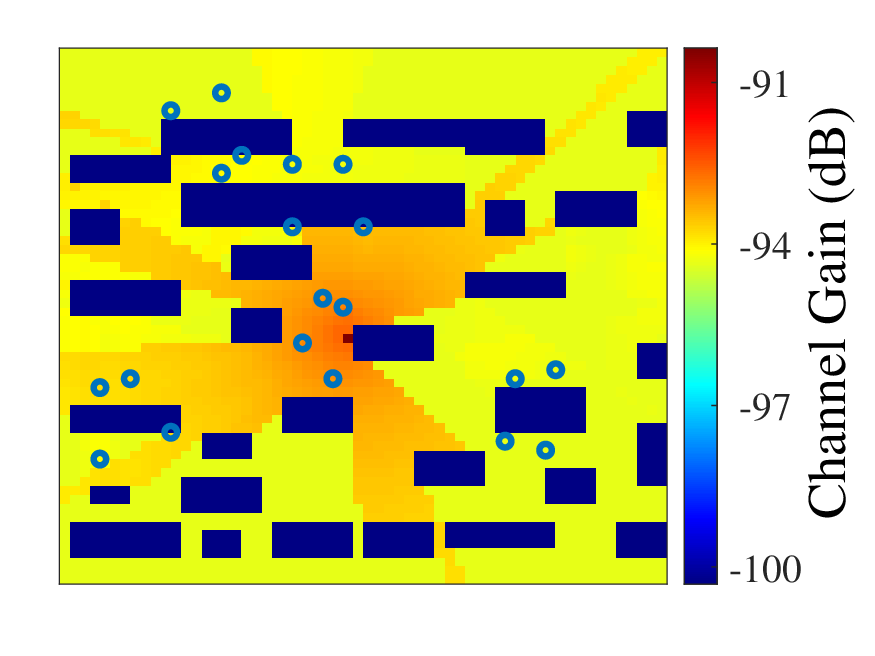}
    \caption{Proposed I, 0.67 }
  \end{subfigure} 
  \hfill
  \begin{subfigure}[b]{0.15\textwidth}
    \centering
    \includegraphics[width=\textwidth]{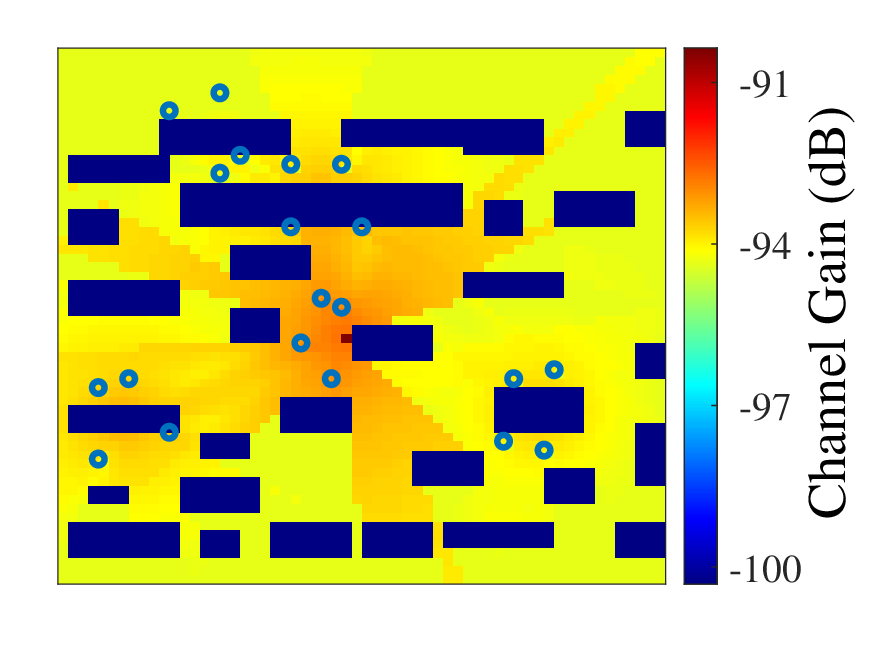}
    \caption{KPSM I, 2.16 }
  \end{subfigure} 
 \hfill
  \begin{subfigure}[b]{0.15\textwidth}
    \centering
    \includegraphics[width=\textwidth]{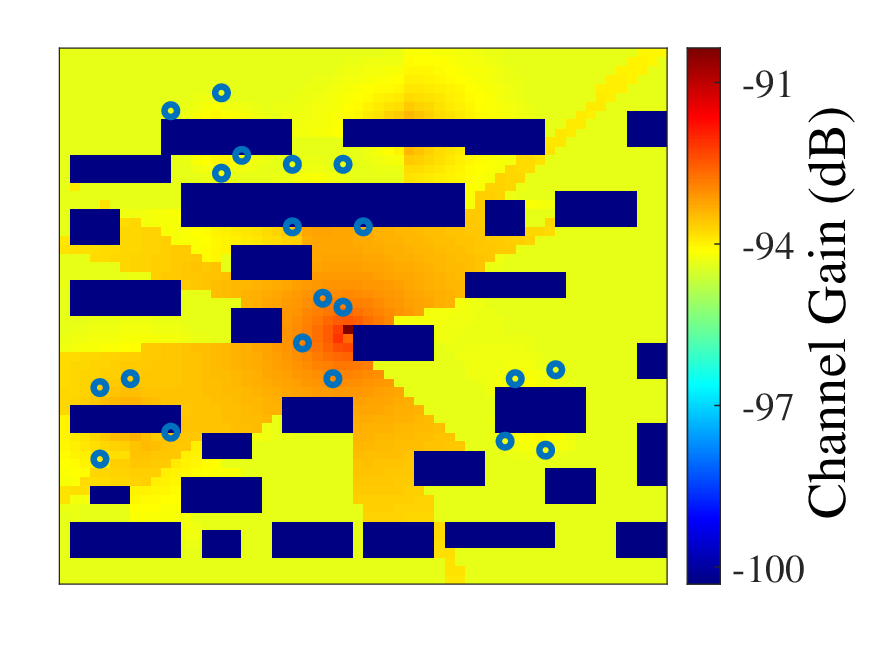}
    \caption{ISSM I, 3.60 }
  \end{subfigure} 
  \hfill
  \begin{subfigure}[b]{0.15\textwidth}
    \centering
    \includegraphics[width=\textwidth]{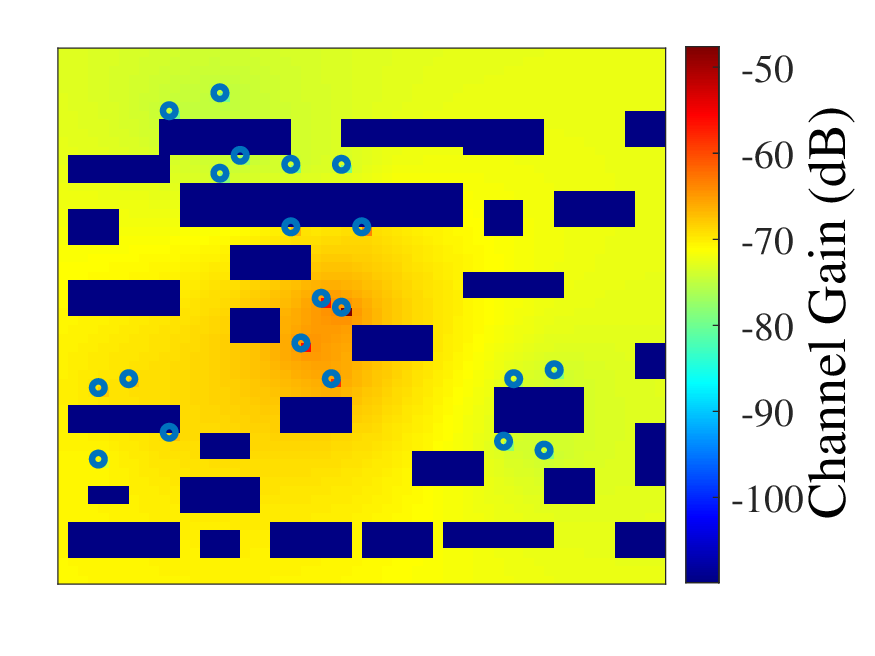}
    \caption{Kriging I, 10.96 } 
      \end{subfigure} 
    \hfill
  \begin{subfigure}[b]{0.15\textwidth}
    \centering
    \includegraphics[width=\textwidth]{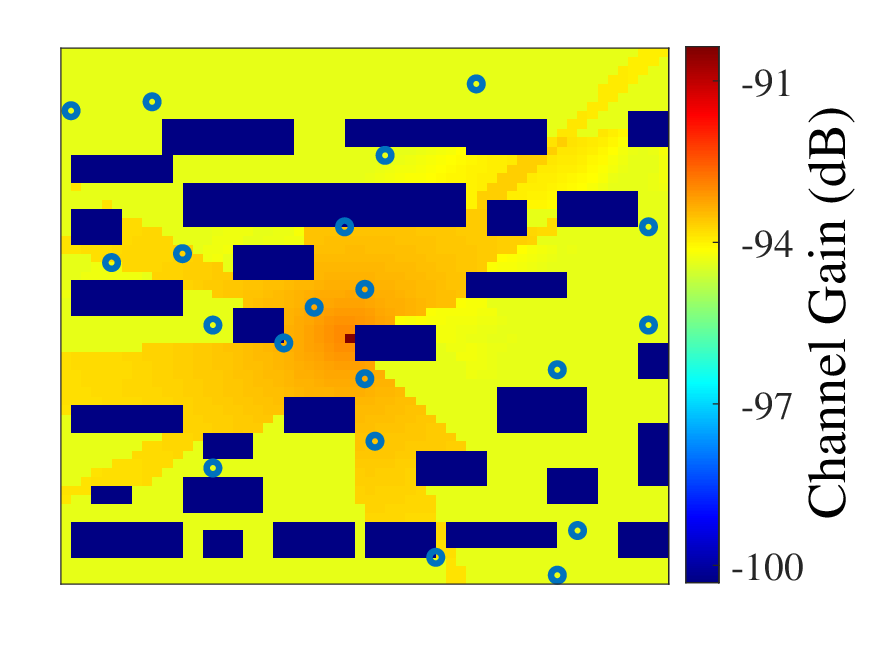}
    \caption{Proposed II, 0.42}
  \end{subfigure}
  \hfill 
  \begin{subfigure}[b]{0.15\textwidth}
    \centering
    \includegraphics[width=\textwidth]{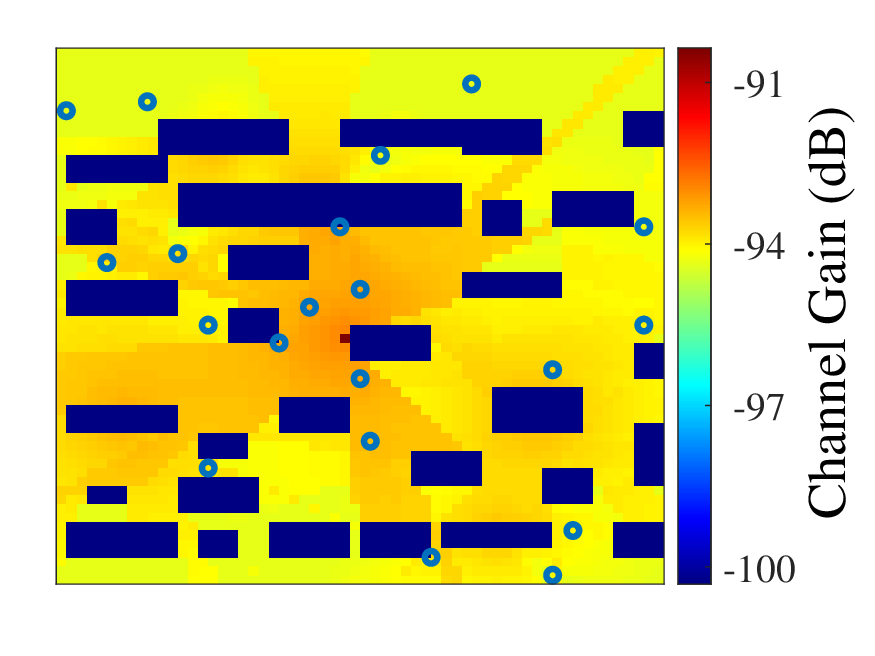}
    \caption{KPSM II, 5.92 }
  \end{subfigure}
  \hfill
  \begin{subfigure}[b]{0.15\textwidth}
    \centering
    \includegraphics[width=\textwidth]{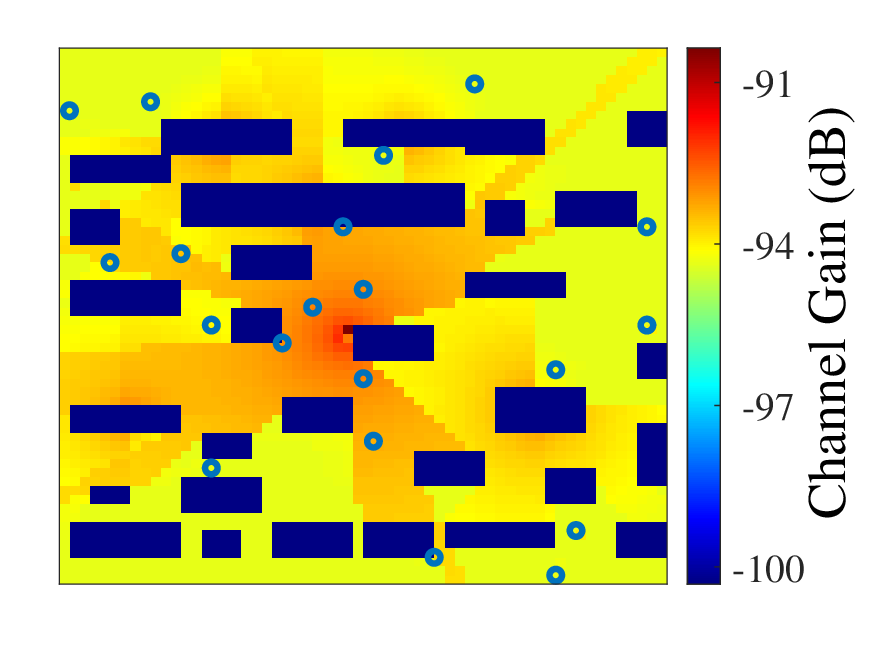}
    \caption{ISSM II, 6.01 }
  \end{subfigure}
  \hfill
  \begin{subfigure}[b]{0.15\textwidth}
    \centering
    \includegraphics[width=\textwidth]{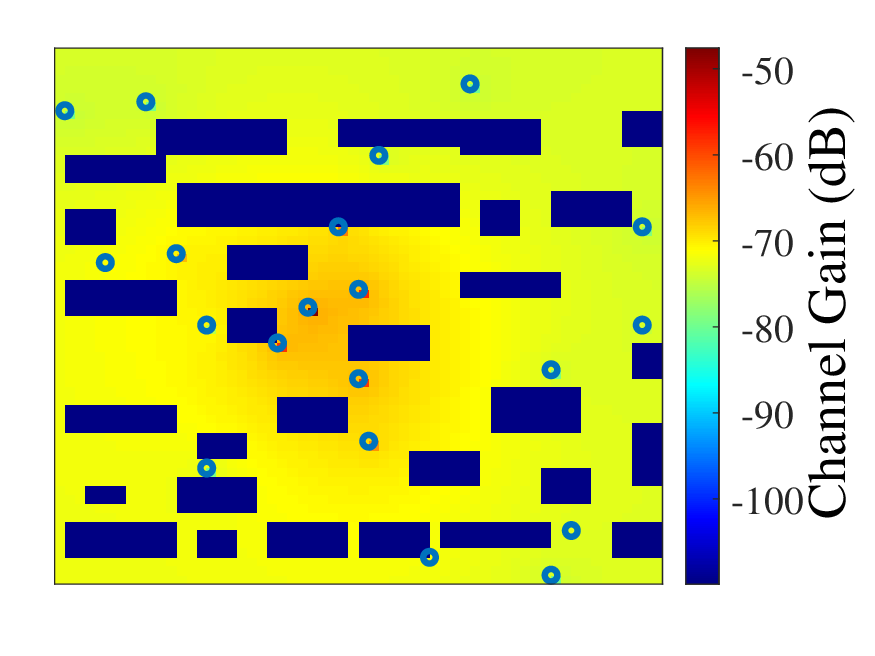}
    \caption{Kriging II, 7.65 }
  \end{subfigure}
  \caption{Ground-truth and reconstructed CGMs.} \vspace{ -16pt } 
  \label{Fig005}
\end{figure}

As shown in Fig. \ref{Fig004}, the simulation setup is generated in a real-world region of size $300$ m $\times$ $300$ m based on ray tracing. In the region, $30$ physical scatterers with different sizes are distributed, resulting in a multipath propagation environment. A Tx is located at the center of the region at a height of $9$ m. The system is operated at the frequency of $2.6$ GHz, such that the wavelength is $\lambda = 0.11$ m. To facilitate CGM characterization, the region is uniformly divided into $60 \times 60$ grids. The side length of each grid is set to $43\lambda$\cite{CGM01}. To evaluate the effectiveness of the proposed algorithms, three CGM construction methods are adopted as benchmarks, detailed as follows: (1) CGM estimation method based on the Kernel-based Physical Scatterer Model (KPSM), in which the SRCs are modeled by the proposed kernel-based SRC model, while the scatterer positions are fixed at the centers of dominant physical scatterers, (2) CGM estimation method with the Independent-Sector-based Scatterer Model (ISSM)\cite{CGM01}, where each physical scatterer is represented as a point with a fixed position, as well as constant and uncorrelated SRCs, and (3) the Kriging method\cite{Kriging}. To ensure a fair comparison, we evaluate the NMSE of each method under two designs for the selection of measurement grids: (i) constrained selection, where grids close to physical scatterers are selected in each sector, as specified in \cite{CGM01}, referred to as ``Type-I'' measurement, and (ii) random selection from $\cal{I}$, referred to as ``Type-II'' measurement. 

Fig. \ref{Fig005} shows the ground-truth CGM and the CGMs constructed using different methods based on $L=20$ measurements, where the locations of the scatterers are indicated in dark blue. The number following the algorithm name in each subfigure title indicates the NMSE values of this algorithm. It can be observed that the proposed virtual scatterer model-based CGM reconstruction method achieves the lowest NMSE among different methods under both Type-I and Type-II measurements. The proposed method outperforms KPSM, which uses a kernel-based SRC model but fixed scatterer number and positions. The accuracy improvement is due to the refinement of scatterer positions by the proposed virtual scatterer model-based CGM construction, which provides more DoFs for CGM estimation than the traditional physical scatterer model with fixed scatterer positions. Nevertheless, compared to the ISSM as shown in Figs. \ref{Fig005}(d) and (h), the CGMs constructed with the KPSM as shown in Figs. \ref{Fig005}(c) and (g) still attain higher accuracy. The reason is that the ISSM method requires at least one measurement grid per AoD sector for each scatterer, leading to a low angular resolution of the scatterer model under a given measurement budget. In contrast, the KPSM permits a finer angular resolution by using the proposed kernel-based SRC model, thus achieving higher CGM estimation accuracy. 

\begin{figure}[t]
  \centering
  \begin{subfigure}[b]{0.215\textwidth}
    \centering
    \includegraphics[width=\textwidth]{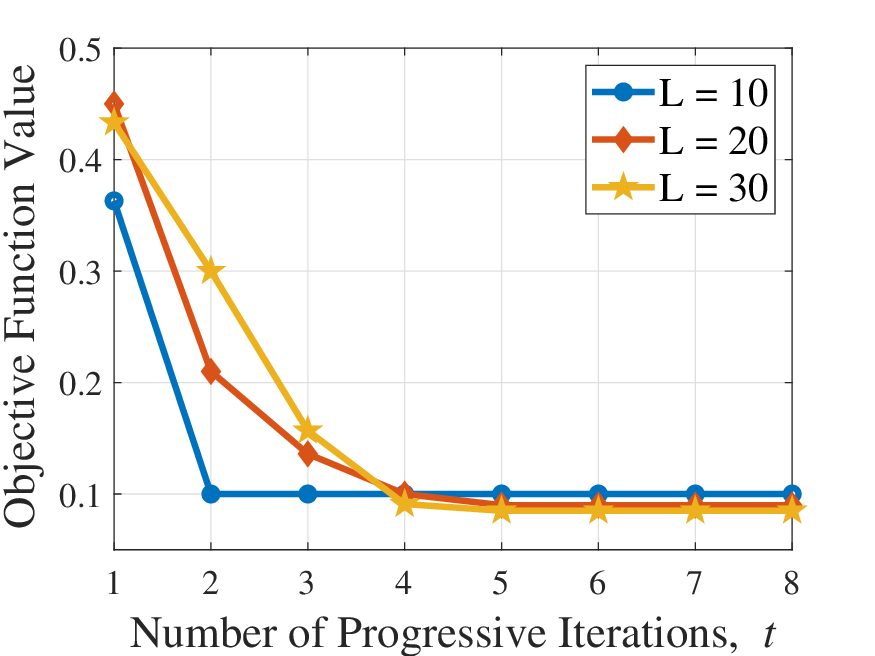} 
  \end{subfigure}
  \hfill
  \begin{subfigure}[b]{0.215\textwidth}
    \centering
    \includegraphics[width=\textwidth]{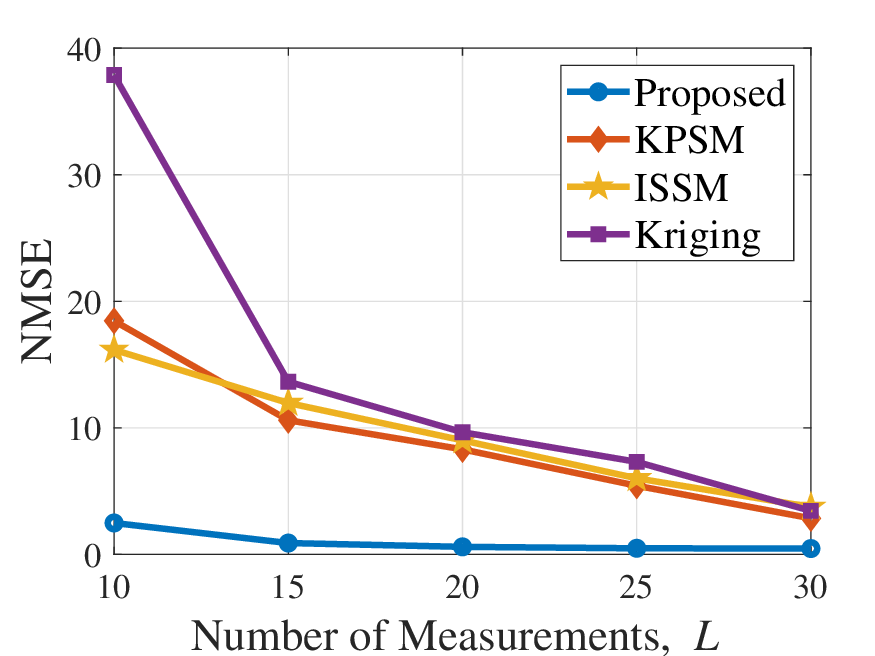} 
  \end{subfigure} 
  \caption{(a) Convergence of the progressive estimation method, (b) NMSE performance versus the number of measurements, $L$.} \vspace{-16pt} 
  \label{fig009}
\end{figure} 
Finally, we investigate the convergence and the NMSE performance of the proposed CGM construction algorithm under different numbers of measurements, $L$. Fig. \ref{fig009}(a) shows that the progressive estimation error decreases and converges as the number of virtual scatterers increases. In particular, the case with $L=10$ measurements converges faster than that with $L=20$ or $L=30$. This is because when only a small number of channel power gain measurements are available, fewer virtual scatterers are sufficient to achieve convergence, as the limited measurements can only support the estimation of a small set of model parameters. Although using fewer measurements leads to faster convergence, it also causes a larger CGM estimation error. As shown in Fig. \ref{fig009}(b), progressive estimation with a larger number of measurements achieves a lower converged NMSE. In addition, the proposed algorithm consistently outperforms the benchmark approaches in terms of NMSE, especially when only a small number of channel measurements are available.

\section{Conclusion} \label{sec005}

 This paper proposed a novel virtual scatterer model with tunable scatterer parameters to characterize the impact of the physical environment on wireless signal propagation. Based on the virtual scatterer model, a progressive scatterer parameter estimation method was developed to reconstruct the CGM based on channel power gain measurements. Moreover, by exploiting the angular correlation of SRCs, a GPR-based inference method was developed to predict the SRCs that cannot be directly estimated but are required for CGM reconstruction, thereby enabling global estimation of channel gains for all grids across the entire region. Simulation results under the 3D ray-tracing channel model validated the effectiveness of the proposed virtual scatterer modeling and CGM estimation approaches. Robustness study on imperfect LoS information, sensitivity analyses, and extensions of virtual-scatterer modeling to general channel map reconstruction (beyond CGM) are worthy of further investigation.  

 \section{Acknowledge} 
This work is supported by National University of Singapore under Research Grants A-8003646-00-00 and A-8003676-00-00. The work is also supported in part by the National Natural Science Foundation of China under grants Nos. 62471424, 92267202, and U25A20390, and the Shenzhen Fundamental Research Program under grant No. JCYJ20250604141209012.

\setlength{\baselineskip}{0.9\baselineskip}
\bibliographystyle{IEEEtran}
\bibliography{mybib}

\end{document}